\documentclass[12pt,a4paper]{article}

\usepackage{amsmath,amssymb}
\usepackage[bookmarks=true]{hyperref}
\usepackage[nosort]{cite}
\usepackage[bulletsep]{collect}
\def\gfxon{\usepackage[final]{graphicx}}

\gfxon

\makeatletter
\let\old@startsection=\@startsection
\renewcommand{\@startsection}[6]{\old@startsection{#1}{#2}{#3}{#4}{#5}{#6\mathversion{bold}}}
\makeatother

\def\eeq{\end{eqnarray}}

\def\=:{=\hspace{-.7em}\raisebox{1.1ex}{.}\hspace{.1em}\raisebox{-0.2ex}{.} }

\newcommand{\beqn}{\begin{eqnarray}}
\newcommand{\eeqn}{\end{eqnarray}}

\newcommand {\beq}{\begin{eqnarray}}
\newcommand {\eeqq}{\end{eqnarray}}

\makeatletter \@addtoreset{equation}{section} \makeatother

\makeatletter
\let\old@makecaption=\@makecaption
\def\@makecaption{\small\old@makecaption}
\makeatother

\makeatletter
\def\mr@ignsp#1 {\ifx\:#1\@empty\else #1\expandafter\mr@ignsp\fi}%
\newcommand{\multiref}[1]{\begingroup
\xdef\mr@no@sparg{\expandafter\mr@ignsp#1 \: }%
\def\mr@comma{}%
\@for\mr@refs:=\mr@no@sparg\do{\mr@comma\def\mr@comma{,}\ref{\mr@refs}}%
\endgroup}
\makeatother

\ifx\href\asklfhas\newcommand{\href}[2]{#2}\fi

\begin{document}

\begin{flushright}\footnotesize
\texttt{FTPI-MINN-11/18} \\
\texttt{UMN-TH-3008/11}
\vspace{0.5cm}
\end{flushright}
\vspace{0.3cm}

\renewcommand{\thefootnote}{\arabic{footnote}}
\setcounter{footnote}{0}
\begin{center}%
{\Large\textbf{\mathversion{bold}
Decomposing Instantons in Two Dimensions}
\par}

\vspace{1cm}%

\textsc{Muneto Nitta$^{1}$ and  Walter Vinci$^{2}$}

\vspace{10mm}
$^1$\textit{Department of Physics, and Research and Education Center 
for Natural Sciences,\\ Keio University, 4-1-1 Hiyoshi, Yokohama, 
Kanagawa 223-8521, Japan}
\\
\vspace{.3cm}
$^2$\textit{University of Minnesota, School of Physics and Astronomy\\%
116 Church Street S.E. Minneapolis, MN 55455, USA}

\vspace{7mm}

\thispagestyle{empty}

\texttt{nitta(at)phys-h.keio.ac.jp}\\
\texttt{vinci(at)physics.umn.edu} 

\par\vspace{1cm}

\vfill

\textbf{Abstract}\vspace{5mm}

\begin{minipage}{12.7cm}

We study BPS vortices in the 1+1 dimensional $\mathcal N=(2,2)$ supersymmetric $U(1)$ gauged  $\mathbb CP^{1}$ non-linear sigma model. We use the moduli matrix approach to analytically construct  the moduli space of solutions and  solve numerically the BPS equations. We identify two topologically inequivalent types of magnetic vortices, which we call S and N vortices. Moreover we discuss their relation to instantons (lumps)  present in  the un-gauged case. In particular, we describe how a lump is split into a couple of component S-N vortices after gauging. We extend this analysis to the case of the extended Abelian Higgs model with two flavors, which is known to admit semi-local vortices.  After  gauging the relative phase between fields, semi-local vortices are also split into component vortices. 

We discuss interesting applications of this simple set-up. Firstly,  the gauging of non-linear sigma models reveals a semiclassical ``partonic'' nature of instantons in 1+1 dimensions. Secondly, weak gauging  provides for a new interesting regularization of the metric of semi-local vortices.

\end{minipage}

\vspace{3mm}

\vspace*{\fill}

\end{center}

\newpage

\section{Introduction}

Instantons play an important  role in quantum field theories 
in various dimensions. 
In four dimensions, they play a prominent role in defining the properties of the QCD vacuum, and in particular in explaining strong coupling effects like chiral symmetry breaking. In the case of ${\cal N}=2$ supersymmetric QCD, they give non-perturbative corrections 
to the exact low-energy effective action
  \cite{Seiberg:1994rs,Seiberg:1994aj}.
It has been long pointed out 
that there are similarities between 
four dimensional gauge theories 
and two dimensional non-linear sigma models, 
such as dynamical generation of a mass gap and asymptotic freedom. These similarities extend to the quantitative level in the supersymmetric case, where it happens that the exact BPS spectra of super QCD and non-linear sigma models are the same \cite{Dorey:1999zk,Dorey:1998yh,Bolokhov:2011mp}. 
It is not a coincidence that 
instantons were found in the $O(3)$ sigma model 
\cite{Polyakov:1975yp} 
almost at the same time of  
the discovery of 
instantons in Yang-Mills theory \cite{Belavin:1975fg}.
The $O(3)$ sigma model is equivalent to 
${\mathbb C}P^1 \,(\simeq SU(2)/U(1))$ sigma model 
and therefore instanton solutions can be extended to 
${\mathbb C}P^{N-1} \,(\simeq SU(N)/[SU(N-1)\times U(1)])$ model. 
In more general terms, when $\pi_2(M)$ is nontrivial for 
the target space $M$, the model admits instantons. 
Instantons in both Yang-Mills theory and sigma models 
have a scale modulus in addition to orientational moduli 
in the internal space and position moduli. In particular, the moduli space of instantons in  four dimensional Yang-Mills has real dimension $4N$, while that of two dimensional sigma model instantons is $2N$.
Sigma model instantons are particle-like object in 
2+1 dimensions, and they are called lumps in field theory \cite{Manton:2004tk}, 
and skyrmions or coreless vortices 
in condensed matter physics. 
The ${\mathbb C}P^1$ model can be lifted to 
a $U(1)$ gauge theory with two complex scalar fields with equal charges, 
which reduces to the ${\mathbb C}P^1$ model in the limit 
of gauge coupling sent to infinity.
The sigma model instanton is lifted to a semi-local vortex \cite{Vachaspati:1991dz,Achucarro:1999it,Hindmarsh:1991jq,Hindmarsh:1992yy,Preskill:1992bf}, 
having the same moduli including a size modulus.

The moduli space dimension of four and two dimensional instantons, together with other circumstantial observations, has led to the conjecture that instantons can be more conveniently thought as being formed of $N$ component ``partons''. The moduli space parameters then describe the positions of these fundamental objects in the euclidean space. The idea that the functional integral of strongly coupled Yang Mills is dominated by a liquid of partons which forms the vacuum, explains many aspects of confinement  and chiral symmetry breaking \cite{Belavin:1979fb,Diakonov:1999ae,Son:2001jm,Zhitnitsky:2006sr}.   
The decomposition of sigma model instantons has been considered also more recently \cite{Collie:2009iz,Eto:2009bz} and it  
can occur by deforming the metric of the target space of 
sigma models. The energy density of the configuration 
has then subpeaks which can then be interpreted as partons. 
The proposal of Ref. \cite{Collie:2009iz} is to identify  these component partons as the UV degrees of freedom which may render 2+1 sigma models and 4+1 gauge theories renormalizable. 
In this paper we discuss, at the semiclassical level, the decomposition of instantons 
in a manner different from \cite{Collie:2009iz,Eto:2009bz,Gudnason:2010yy}.
To this end,  we consider a two dimensional  $\mathcal N=(2,2)$ supersymmetric  $\mathbb CP^{1}$ non-linear sigma model. We then gauge a $U(1)$ subgroup of the full $SU(2)$ isometry enjoyed 
by the ${\mathbb C}P^1$ model. 
The $U(1)$ gauged ${\mathbb C}P^1$ model,  
is known to admit two Abelian vortices 
(Abrikosov-Nielsen-Olesen vortices \cite{Abrikosov:1956sx,Nielsen:1973cs})
\cite{Schroers:1995ns,Schroers:1996zy,Baptista:2004rk,Baptista:2005bt,Baptista:2010rv}.
We show how an instanton is decomposed into these vortices.

\begin{figure}[htbp]
\begin{center}
\includegraphics[width=10cm]{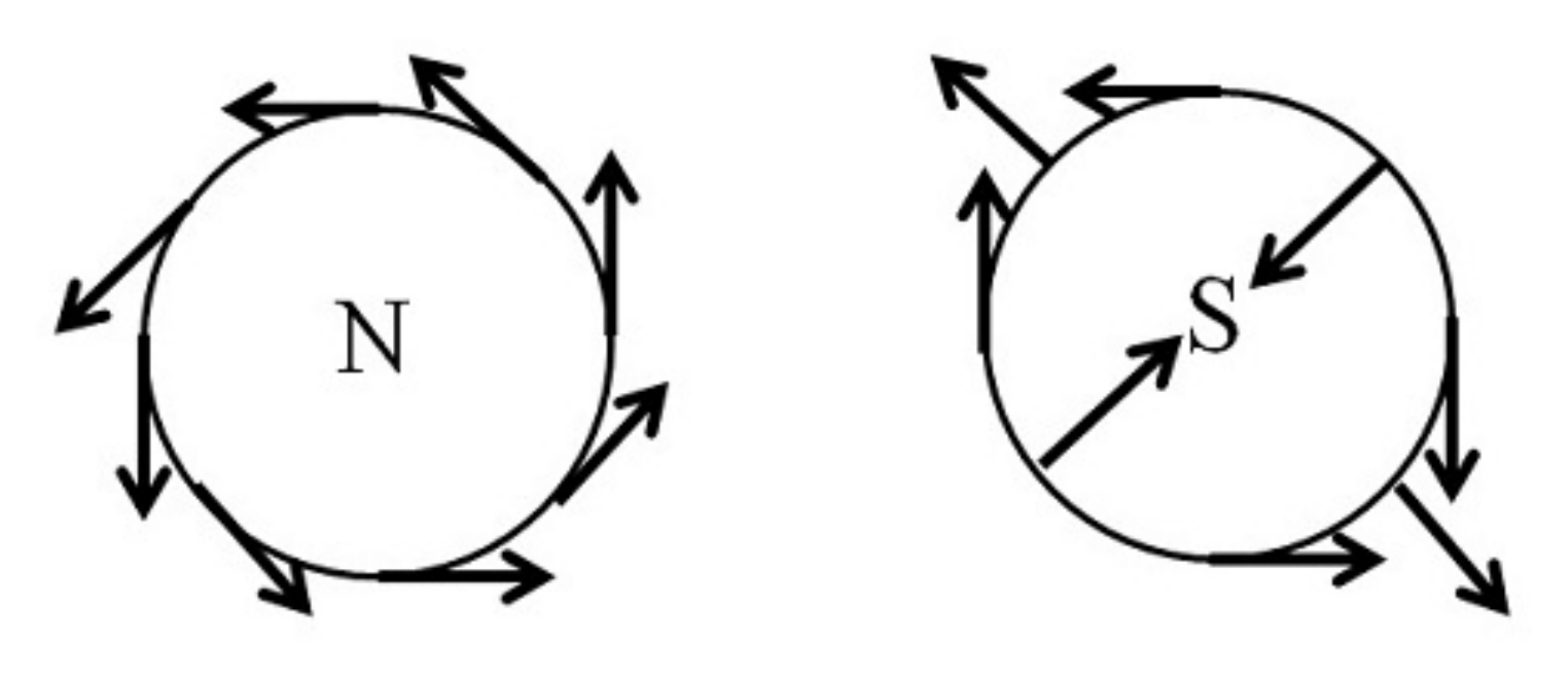}\caption{
N and S vortices. 
The boundary conditions of the N and S vortices are shown by 
arrows which represent points on the gauge orbit 
(on the equator) of $S^2$.
N and S denote the north and south poles in the vortex cores.}
\label{fig:NSvortices}
\end{center}
\end{figure}

\begin{figure}[htbp]
\begin{center}
\includegraphics[width=10cm]{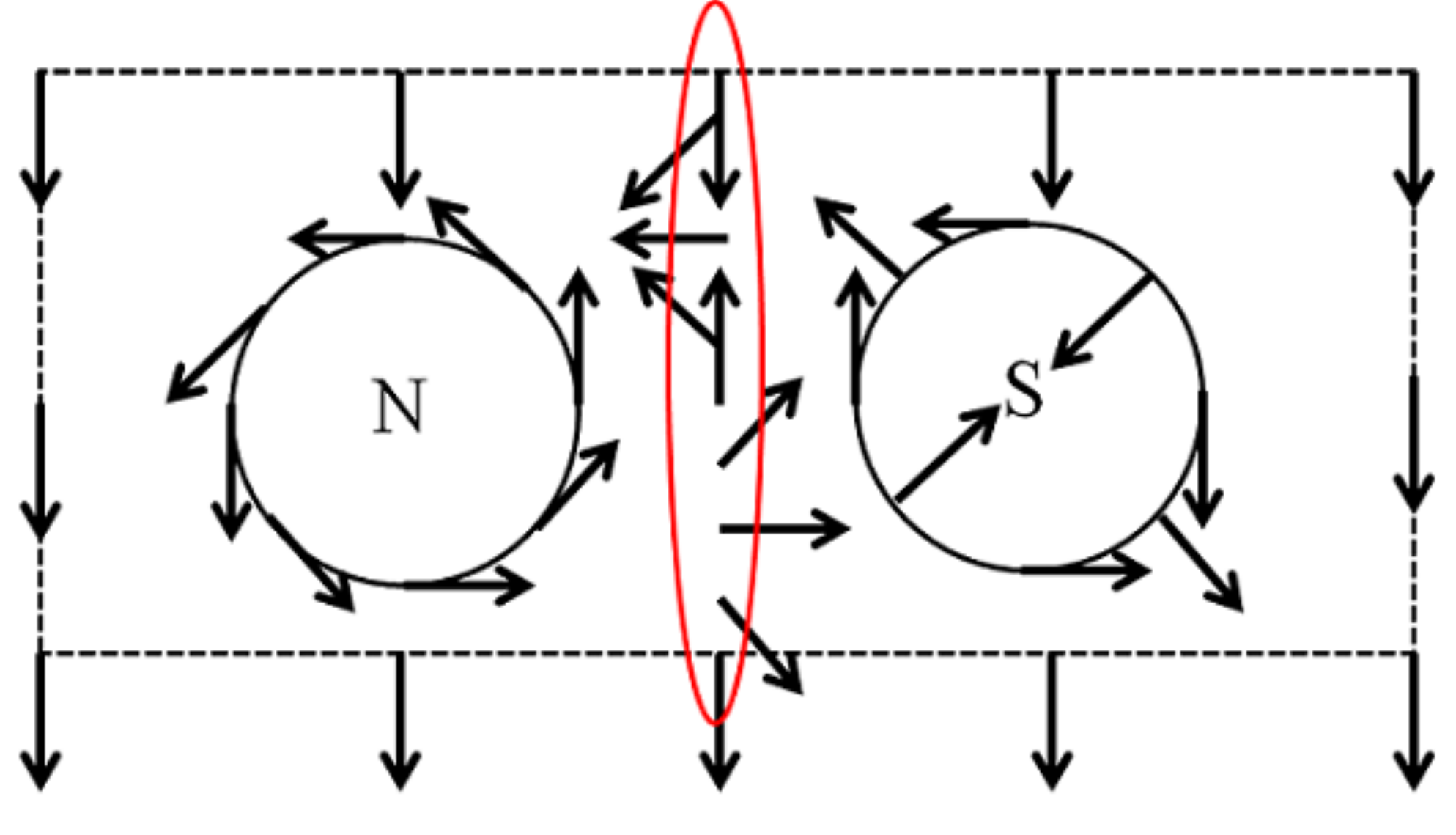}\caption{The decomposition of a 
lump into N and S vortices. 
The whole configuration is a lump with charge one,
where the arrows represent points on $S^2$. 
Configurations of N and S vortices appear.
A gauge orbit exists between them (around the oval region), 
and it separates the N and S vortices.
}
\label{fig:lump}
\end{center}
\end{figure}

The supersymmetric gauged ${\mathbb C}P^1$ model  
admits a unique supersymmetric vacuum 
up to the gauge symmetry, while the vacuum manifold 
is a $U(1)$ gauge orbit, a circle, on $S^2$.
The potential, completely fixed by supersymmetry, has two local maxima at 
the north and south poles. 
Encircling the outside of a vortex core at the boundary, 
the $U(1)$ phase winds once around the gauge orbit.
When this winding is unwound at the core of a vortex, 
there exists two choices, the north and south poles, 
see Fig.~\ref{fig:NSvortices}.
Therefore there exist two kinds of vortices with different cores.
We call them N and S vortices in this paper.
One instanton wraps $S^2$ once while N and S vortices 
wrap upper and lower hemispheres 
bounded by the vacuum $U(1)$ gauge orbit, see Fig.~\ref{fig:lump}.
Accordingly, each of them has a half (or more generally fractional) 
instanton charge of $\pi_2(M)$, 
so that they can be called two-dimensional merons \cite{Gross:1977wu},
in analogy with merons in four dimensions
\cite{Callan:1977qs,Callan:1978bm}.
A set of N and S vortices can be interpreted as one instanton 
as can be seen in Fig.~\ref{fig:lump},  
where the distance between them corresponds to 
the size of the instanton. 
In this sense N and S vortices are constituents 
of an instanton. 
It was found that the moduli space metric is incomplete 
at the point where the positions of N and S vortices coincide 
\cite{Baptista:2004rk}. 
In our understanding, this is nothing but a small instanton 
singularity.

Once the ${\mathbb C}P^1$ model is constructed as the low energy limit of  
a linear $U(1)$ gauge theory, 
the $U(1)$ gauged ${\mathbb C}P^1$ model can be formulated 
as $U(1) \times U(1)$ gauge theory with two complex scalar fields. In this way, the moduli space of instantons is promoted to that  of semi-local vortices, which is regular.
Then, the small instanton singularities are resolved.
No pathology occurs when N and S vortices coincide, 
so that the moduli space is regularized. 

Another advantage in considering this $U(1) \times U(1)$ linear formulation is that we can obtain a rather interesting regularization of the semi-local vortex metric. An important recent development concerning the aforementioned correspondence between BPS spectra in two and four dimensions  regards the role played by non-Abelian semi-local vortices. More precisely, the correspondence holds between four dimensional $U(N)$ super QCD with $N_{f}>N$ flavors and the two dimensional effective theory on the vortex world-sheet \cite{Hanany:2004ea,Shifman:2004dr} hosted by the theory when put on the Higgs phase. This correspondence has been proved using D-brane constructions of the vortex theory. However, while string theory gives a well-defined effective theory, it is well known that some zero-modes are non-normalizable when the effective theory is derived from field theory \cite{Shifman:2006kd,Eto:2007yv}. It is then difficult to  quantize the vortex theory and check the correspondence in a fully field theoretic framework. A possible approach to the problem has been considered for example in Ref. \cite{Koroteev:2011rb}. Here we propose weak gauging as an alternative approach to the problem. The weak gauging of a non-linear sigma model, when considered as the effective theory of a semi-local vortex string, should correspond to a deformation of the four dimensional bulk theory. This would give  new 4d/2d correspondences together with important insights on the physics of non-normalizable modes in the undeformed case.

While we have considered only the supersymmetric case, we can also use our analysis to obtain qualitative answers about non-supersymmetric cases. A more generic potential would introduce for example some non-trivial interactions between the component vortices. There exist  examples of analogs of 
these non supersymmetric vortices in condensed matter systems, 
in which also one or both of the $U(1)$ symmetries are 
 global. The case when both $U(1)$s are global is relevant in the case of anti-ferromagnets and 
two component Bose-Einstein condensates of ultracold atoms 
in the anti-ferromagnetic phase \cite{Kasamatsu:2005,Eto:2011wp}.
Even if these systems have a global symmetry differing from 
the supersymmetric gauged ${\mathbb C}P^1$ model,
 they have similar potential terms. 
Consequently topological properties, 
for instance whether and how instantons 
are decomposed into constituent vortices, are the same.
Another condensed matter example which is similar to our model 
is given by two-band superconductors.
The Landau-Ginzburg Lagrangian proposed to describe them 
consists of two gaps (complex scalar fields) 
coupled to the electro-magnetic $U(1)$ gauge field.
It has $U(1) \times U(1)$ symmetry, one of which 
is local electro-magnetic $U(1)$ symmetry while 
the other $U(1)$ symmetry, a relative phase between 
the two fields, is a global $U(1)$ which is 
broken explicitly in the presence of Josephson interactions 
between two gaps. 
This model is also known to admit fractional vortices 
\cite{Babaev:2001hv,Babaev:2002ck}
which are mostly the same with ours 
in the absence of Josephson interactions\footnote{
However in the presence of Josephson interactions 
two kinds of vortices are connected by a domain wall 
\cite{Tanaka:2001a,Tanaka:2001b,Goryo:2007}.
}. 
Study of our supersymmetric gauged ${\mathbb C}P^1$ model 
will give insights to these condensed matter systems.

The structure of the paper is the following. In Section 2 we briefly review the construction of two-dimensional instantons as $\mathbb CP^{1}$ lumps. We then employ the moduli matrix formalism to construct vortices in the gauged version of the sigma model. In Section 3 we construct a bound state of N and S vortices and show how they emerge from an instanton configuration as we increase the strength of the gauge interactions. In Section 4 we lift the $\mathbb CP^{1}$ non-linear sigma model to a $U(1) \times U(1)$ linear formulation and consider regularization of small instanton singularities. In Section 5 we discuss how the gauging of flavor symmetries can provide for a nice regularization of the semi-local vortex metric. Finally, in Section 6 we present various possible generalizations where we consider sigma models with higher dimensional target spaces and gauging of flavor symmetries of higher rank.


\section{Solitons in Gauged $\mathbb CP^{1}$ non-Linear Sigma Model}
\label{sec:gauged}

\subsection{Models}
\subsubsection*{Un-gauged $\mathbb CP^{1}$}

Let us start by considering the standard two-dimensional $\mathbb CP^{1}$ non-linear sigma model (NL$\sigma$M).  The action is the following \cite{Polyakov:1975yp}:
\beq
\mathcal L = \xi \frac{|\partial_{\mu} b|^{2}}{(1+|b|^{2})^{2}}\,,\quad \mu=1,2. 
\label{eq:ungauged}
\eeq
In the formula above, $b$ is the holomorphic coordinate parameterizing the target space, and we work with the euclidean metric. The metric for $b$ is given by the standard round Fubini-Study metric and it  is written in a patch which does not include the point at infinity $b=\infty$. We can include this point performing a  change of variable $b=1/b'$, under which the lagrangian (\ref{eq:ungauged}) is indeed invariant.   The standard Fubini-Study metric  is of the K\"ahler type, since it can be derived from a K\"ahler potential:
\beq
\mathcal L =\partial_{b}\partial_{\bar b} \, K(b,\bar b)\,|\partial_{\mu} b|^{2}, \quad K=\xi \log (1+ |b|^{2})\,.
\eeq
The K\"ahler property of the metric ensures the existence of a supersymmetric extension with four supercharges of the bosonic NL$\sigma$M \cite{Zumino:1979et,AlvarezGaume:1981hm}. The analyses of this paper can be extended to a study of static string-like solutions with translational symmetry in 3+1 dimensions. Formula (\ref{eq:ungauged}) will then represent the dimensionally reduced four-dimensional action to  the plane transverse to the string.
From the point of view of constructing solitons, the model we consider can be considered as either being a truncated  bosonic sector of an  $\mathcal N=2$ theory in four dimensions or an $\mathcal N=(2,2)$ theory in two dimensions. Since our analysis is limited to the classical level, we never show explicitly the fermionic sector. Supersymmetry however fixes  the kinetic terms and the potentials we will consider later.

Since the  $\pi_{2}$ of the target space is nontrivial, the  $\mathbb CP^{1}$ NL$\sigma$M contains stable topological solitons of codimension two. This means string-like objects in four dimensions or instantons in two euclidean dimensions. Solitons can be found with a square root completion of the Bogomol'nyi type\cite{Polyakov:1975yp}.  If we restrict ourselves to consider  static solutions, we can rewrite the lagrangian (\ref{eq:ungauged}) in the following way 
\beq
\mathcal E&  =&  2 \,\xi \frac{|\partial_{z} b|^{2}+|\partial_{\bar z} b|^{2}}{(1+|b|^{2})^{2}}= 2 \pi \xi N + 4 \,\xi \frac{|\partial_{\bar z} b|^{2}}{(1+|b|^{2})^{2}}\,, \nonumber \\
z& \equiv& x_{1}+i\,x_{2}\, \quad   \partial_{z}\equiv\frac12(\partial_{1}-i \partial_{2})\,.
\label{eq:ungaugedcomp}
\eeq
The  quantity
\beq
N \equiv \frac1\pi \frac{|\partial_{z} b|^{2}-|\partial_{\bar z} b|^{2}}{(1+|b|^{2})^{2}}
\label{eq:lumpnumb}
\eeq
 gives the degree of the map $b(z): \mathbb CP^{1} \rightarrow\mathbb CP^{1}$ and is the  topological integer which characterizes the  second homotopy group:
\beq
\pi_{2}(\mathbb CP^{1})=\mathbb Z\,.
\eeq
 Clearly the energy has a lower bound $E\ge2 \pi \xi N$ which is saturated when the following equation is satisfied:
\beq
\partial_{\bar z} b=0\,.
\eeq

Topological solitons of the type above are  BPS (Bogomol'nyi-Prasad-Sommerfeld) saturated \cite{Bogomol'nyi:1975de,Prasad:1975kr}. They satisfy first order equations  of motion obtained from a square root completion and their energy is proportional to a topological integer. In theories with extended supersymmetry, the mass of BPS solitons  saturates the bound given by a central extension of the supersymmetry algebra \cite{Witten:1978mh}. In the language of supersymmetry, lumps are 1/2 BPS, in the sense that they preserve 1/2 of the supersymmetry transformations \cite{Edelstein:1993bb}. Fermions and supersymmetry are crucial to preserve BPS saturation once quantum effects are taken into account \cite{Rebhan:2003bu}.

Since the energy is proportional to the topological integer $N$, which also counts the number of solitons, there are no static interactions among lumps. The consequence of this fact is the existence of a large set of degenerate solutions (moduli space) parameterized by the positions and orientation of the single component lumps. This degeneration can be understood easily from a mathematical point of view if we notice that the equation above implies that $b$ is a holomorphic function of the complex variable $z$. It must contain a finite number of poles and zeroes, and must then be given by a holomorphic rational function \cite{Polyakov:1975yp}.
\beq
b(z)=b_{\infty}\frac{z^{N}+p_{1}^{N-1}+\cdots+p_{N}}{z^{N}+q_{1}^{N-1}+\cdots+q_{N}}\,,\qquad b_{\infty}\equiv b(\infty)\,,
\label{eq:ratmap}
\eeq
where the degree $N$ of the polynomials is the  lump number (\ref{eq:lumpnumb}).  A fundamental lump is given for example by the following choice:
\beq
b_{0}(z) = b_{\infty}\frac{z-z_{S}}{z-z_{N}}.
\label{eq:singlump}
\eeq

To extract physical quantities such as size and position from the rational map  above, we can make use of the $SU(2)$ isometry enjoyed by the  $\mathbb CP^{1}$ NL$\sigma$M  which acts non-linearly on the field $b$:
\beq
b \rightarrow \frac{v+u b}{u^{*}-v^{*}b}\,,\quad  \quad  \left(
\begin{array}{ccc}
 u^{*} & -v^{*}    \\
   v&  u   \\ 
\end{array}
\right)\in SU(2)\,, \quad   \quad |u|^{2}+|v^{2}|=1\,.
\label{eq:symmetry}
\eeq
We can then always set $b_{\infty}\rightarrow 0$. This puts the moduli matrix (\ref{eq:singlump}) in the form\footnote{If the expectation value of $b$ is vanishing, we must have a rational map where the degree of the numerator is less than the degree of the denominator.}:
\beq
b'_{0}=\frac{\rho}{z-z_{0}},
\eeq
which describes a lump of position $z_{0}$ and the size $\rho$. By explicitly performing this rotation we obtain\footnote{Intuitively, the center of the lump is mapped to $b(z_{0})=-1$, the point diametrically opposed to $b_{\infty}=1$. The size is given by the displacement from the trivial map (zero size lump) $z_{S}=z_{N}=0$.}:
\beq
z_{0}=\frac{z_{S}+|b_{\infty}|^{2}z_{N}}{1+|b_{\infty}|^{2}},\quad \rho= \frac{b_{\infty}}{1+|b_{\infty}|^{2}}(z_{S}-z_{N})\,.
\label{eq:posandsize}
\eeq

\subsubsection*{Gauged $\mathbb CP^{1}$}

We now consider the gauged version of the $\mathbb CP^{1}$ NL$\sigma$M obtained from (\ref{eq:ungauged})  by gauging the following $U(1)$ subgroup of the $SU(2)$ isometry (\ref{eq:symmetry})
\beq
U(1): 
\left(
\begin{array}{ccc}
 e^{-i \theta} & 0    \\
  0 & e^{i \theta} \\
  \end{array}
\right), \quad u=e^{-i \theta},\quad v=0\,,
\eeq
which  acts linearly on the field $b$ 
\beq
b \rightarrow e^{-2 i \theta } b.
\eeq
 The  lagrangian we obtain is then  the following \cite{Bagger:1982fn}:
\beq
\mathcal L_{U(1)} =-\frac{1}{4 g^{2}} F_{g}^{\mu\nu}F_{g\mu\nu}+ \xi \frac{|\nabla_{\mu} b|^{2}}{(1+|b|^{2})^{2}}-\frac{g^{2}}{2}\left( \xi\frac{-2|b|^{2}}{1+|b|^{2}}-\zeta  \right)^{2},
\label{eq:gauged}
\eeq
where 
\beq
\nabla_{\mu}=\partial_{\mu}+2 i A_{g\mu}\,.
\eeq
The complicated potential term is the necessary one for the existence of  BPS saturated solitons. It  can be more easily derived as a D-term potential which arises as one imposes supersymmetry\footnote{See the Appendix for more details.}. In the language of supersymmetry, $\zeta$ is a Fayet-Iliopoulos term \cite{Fayet:1974jb}. 

The existence of  BPS solitons requires that the potential term vanishes in the vacuum of the theory\footnote{This is related to the existence of a supersymmetric vacuum in the supersymmetric extension of the model}. This occurs for
\beq
|b_{\infty}|^{2}=-\frac{\zeta}{\zeta+2 \xi}\,.
\label{eq:vaceq}
\eeq
The Fayet-Iliopoulos term must then assume values within the following range:
\beq
-2 \xi \le \zeta \le 0\,.
\eeq 
In a generic vacuum the expectation value of $b$ spontaneously breaks the $U(1)$ gauge symmetry, and the model contains ANO (Abrikosov-Nielsen-Olesen) vortices  \cite{Abrikosov:1956sx,Nielsen:1973cs} supported by the following homotopy group:
\beq
\pi_{1}(U(1))=\mathbb Z\,.
\label{eq:vortexnaive}
\eeq

Notice that all the discussions of this Section have been carried out considering only one coordinate patch for the $\mathbb CP^{1}$ target space. As already mentioned, to cover the full manifold we need to take into account another  coordinate patch obtained with the change of variable:
\beq
b'= 1/b\,.
\eeq
In the gauged case, after this change of variables the lagrangian (\ref{eq:gauged})  takes the form
\beq
\mathcal L_{U(1)} =-\frac{1}{4 g^{2}} F_{g}^{\mu\nu}F_{g\mu\nu}+ \xi \frac{|\nabla_{\mu} b'|^{2}}{(1+|b'|^{2})^{2}}-\frac{g^{2}}{2}\left( \xi\frac{-2|b'|^{2}}{1+|b'|^{2}}-\zeta'  \right)^{2},
\eeq
which has the same form of (\ref{eq:gauged}) provided that  the  FI term $\zeta$ transforms nontrivially while we change coordinate patch: 
\beq
\zeta'=-\zeta-2 \xi\,.
\eeq


\subsection{Vortex equations}
The first step in the study of vortex solutions is to employ a Bogomol'nyi completion of the action (\ref{eq:gauged}) \cite{Bogomol'nyi:1975de}:
\beq
\mathcal E_{U(1)}& =&  \frac{1}{2g^{2}}\left[F_{g12}-g^{2}\left(\xi\frac{-2|b|^{2}}{1+|b|^{2}}-\zeta\right)\right]^{2} + 4 \,\xi \frac{|\nabla_{g \bar z} b|^{2}}{(1+|b|^{2})^{2}}+\nonumber\\[2mm]
& - & \zeta F_{g12}  + \xi\epsilon_{ij}\partial_{i}\mathcal N_{j}. \nonumber \\
\nonumber \\
\mathcal N_{j}& =   &      \frac{i}{2(1+|b|^{2})}\left(  b \,\nabla_{j}\bar b -\bar b \, \nabla_{j} b \right)\,.
\label{eq:bogcompl}
\eeq
As usual, vortex equations are given by imposing the vanishing of the squares in the first line of the expression above:
\begin{eqnarray}
F_{g12} & = & g^{2}\left(\xi \frac{-2|b|^{2}}{1+|b|^{2}} -\zeta\right)\nonumber \\
\nabla_{g \bar z} b& = & 0 \,.
\label{eq:bog}
\end{eqnarray}

The total tension is then given by the two surface terms in the second line. The first term is proportional to the magnetic flux density. The corresponding second term in linear sigma models is usually discarded  as a vanishing boundary term. However, it is non-vanishing for  compact NL$\sigma$M. 
To give the correct topological  interpretation to the two surface terms  above, we have first to discuss in more details the topology supporting vortices in the $\mathbb CP^{1}$ NL$\sigma$M. Vortices are undoubtedly characterized by the homotopy group  (\ref{eq:vortexnaive}). However, when we consider compact spaces,  equation (\ref{eq:vortexnaive}) does not give a complete  classification of vortices \cite{Baptista:2004rk}. We can in fact unwind the circle $b=|b_{\infty}|e^{i\theta}$ representing a vortex configuration at the boundary by either crossing the point $b=0$ (``south pole'') or the point $b=\infty$ (``north pole'').
The topology which correctly describes vortices including core structures 
in the gauged $\mathbb CP^{1}$ sigma model is then the following:
\beq
\pi_{1}(U(1))|_{\mathbb CP^{1}}=\mathbb Z_{S}\times \mathbb Z_{N}\,.
\eeq
We will distinguish vortices with different $\mathbb Z$ charges by labeling  them as S and N vortices.
 The corresponding magnetic fluxes (proportional to topological charges), as we shall prove soon, may then be written in the following way
\beq
\mathcal V_{jS}& =& \mathcal N_{j}\,; \nonumber \\
\mathcal V_{jN}& =& \mathcal N_{j}+2 A_{gj}\,,
\label{eq:topodens}
\eeq
and the energy density can be written as:
\beq
\mathcal E_{U(1)}=   \frac{\zeta+2\xi}{2} \epsilon_{ij}\partial_{i}\mathcal V_{jS}- \frac\zeta2  \epsilon_{ij}\partial_{i}\mathcal V_{jN}\,.
\eeq

\subsubsection*{Moduli matrix}

We now employ the moduli matrix approach \cite{Taubes:1979tm,Gibbons:1992gt,Isozumi:2004vg,Eto:2005yh,Eto:2006pg} to construct vortices and  their moduli space. The moduli matrix construction can be considered as a direct generalization of the rational map construction for lumps to the gauged case \cite{Hindmarsh:1991jq,Vachaspati:1991dz,Eto:2007yv}\footnote{Similar connections between the moduli matrix and the rational map construction for monopoles were discussed in   Ref. \cite{Nitta:2010nd}. }. As usual, we do so by performing  the following change of variables
\beq
b(z,\bar z)  & = & s^{2}(z,\bar z) \,b_{0}(z), \nonumber \\
A_{g \bar z} & = & -i \partial_{\bar z} \log s\,,
\label{eq:def}
\eeq
where $s$ is a non-vanishing function, while $b_{0}$ is holomorphic. Notice that the change of variables above introduces  an un-physical  ``V-equivalence'' which scales  $s$ and $b_{0}$ by multiplication with a constant $V$ 
\beq
s^{2}\rightarrow V^{-1} s^{2},\quad b_{0} \rightarrow V b_{0}
\eeq
but leaves all physical quantities unchanged.   The  second of the equations (\ref{eq:bog}) is then  identically solved, while the first  reduces to a second order gauge invariant master equation:
\beq
\partial_{\bar z}\partial_{z} \log \omega& = & -\frac{g^{2}}{4} \left(\xi\frac{-2\omega^{2}|b_{0}|^{2}}{1+\omega^{2}|b_{0}|^{2}}-\zeta \right)\,, \quad \omega=s s^{\dagger}\,.
\label{eq:master}
\eeq 
The holomorphic function $b(z)$ is called moduli matrix, and its complex coefficients parameterize the moduli space of vortices in the system considered. Let us assume that $b(z)$ has a finite number of zeroes and poles. Then it can be written as a ratio of polynomials
\beq
b_{0}(z)& =& b_{\infty}\frac{z^{n_{S}}+p_{1}^{n_{S}-1}+\cdots+p_{n_{S}}}{z^{n_{N}}+q_{1}^{n_{N}-1}+\cdots+q_{n_{N}}}.
\label{eq:modmatr}
\eeq
In the expression above, the overall coefficient $b_{\infty}$ is not a modulus, since it can be fixed by V-equivalence. We chose it to be equal to the expectation value of $b$. The master equation for $s$ must then be solved (numerically) with the following boundary conditions
\beq
|s|&  \rightarrow & |z|^{n_{N}-n_{S}}  \qquad {\rm as} \qquad  |z| \rightarrow \infty\,.
\eeq
The coefficients $p_{i}$ and $q_{j}$  represent moduli of the vortex configuration, and the complex dimension of the moduli space is:
\beq
{\rm Dim}_{\mathbb C}\, \mathcal M=n_{S}+n_{N}\,.
\eeq

Using the moduli matrix formalism we can now easily evaluate the flux densities (\ref{eq:topodens})
\beq
& & \int dx^{2} (\partial_{1} \mathcal V_{2S}-\partial_{2} \mathcal V_{1S})=2 \int dx^{2} \partial_{z}\partial_{\bar z} \log \left(\frac{1+\omega^{2}|b_{0}|^{2}}{\omega^{2}|b_{0}|^{2}}\right)= 2 \pi n_{S}\,;\nonumber \\
& &\int dx^{2} (\partial_{1} \mathcal V_{2N}-\partial_{2} \mathcal V_{1N})=2 \int dx^{2} \partial_{z}\partial_{\bar z} \log (1+\omega^{2}|b_{0}|^{2})=2 \pi n_{N}\,,
\eeq
which define the integers $n_{S}$ and $n_{N}$ appearing as the degree of the polynomials in  moduli matrix as the S and N-vortex numbers\footnote{Since the function $s$ is non-vanishing, each zero (pole) of $b_{0}(z)$ clearly correspond to the only points where $b$ equals the value at the south or north pole, roughly corresponding to the cores of S and N vortices.}. The total tension is then
\beq
T_{U(1)}=  \pi (\zeta+2\xi) n_{S}-\pi \zeta  n_{N}\,.
\eeq
Notice that the total magnetic flux  is:
\beq
2 \pi \nu_{v}=  -\int dx^{2} F_{g12}=- 2 \int dx^{2} \partial_{z}\partial_{\bar z} \log \omega=\pi (n_{S}-n_{N})\,,
\eeq
thus S and N-vortices have opposite $U(1)$ charges. We can also define a ``fractional lump number''
\beq
\nu_{l}=\frac{n_{S}+n_{N}}{2}\,.
\label{eq:vortnumb}
\eeq
This definition is justified if we notice that the integer above is given, in fact, by the integral of following quantity:
\beq
(\mathcal V_{jS}+\mathcal V_{jN})/2=  \frac{i}{2(1+|b|^{2})}\left(  b \,\nabla_{j}\bar b -\bar b \, \nabla_{j} b \right)-A_{gj}\,,
\eeq
which reduces to the  lump number given in Eq.~(\ref{eq:lumpnumb}) for the un-gauged case when $A_{gj}$ vanishes.
\subsection{S(N) isolated vortex}
In this Section we construct a fundamental vortex and  numerically solve the master equation (\ref{eq:master}). A fundamental S-vortex is given by the following choice for the moduli matrix:
\beq
b_{0}(z)=b_{\infty}(z-z_{S}), \quad |\omega|^{2}\rightarrow |z|^{-1}\,.
\eeq
In fact, the vortex unwinds around the South pole $b=0$ at the center $z_{S}$ of the vortex. Notice from Eq.~(\ref{eq:vortnumb}) that the S-vortex has fractional lump charge $\nu_{l}=1/2$. Similarly, we have an N-vortex with the choice:
\beq
b_{0}(z)=b_{\infty}\frac{1}{z-z_{N}}, \quad |\omega|^{2}\rightarrow |z|\,,
\eeq
where the vortex unwind around the point $b=\infty$ at the point $z_{N}$. The N-vortex also has fractional lump charge $\nu_{l}=1/2$.  

\begin{figure}[htbp]
\begin{center}
\includegraphics[width=10cm]{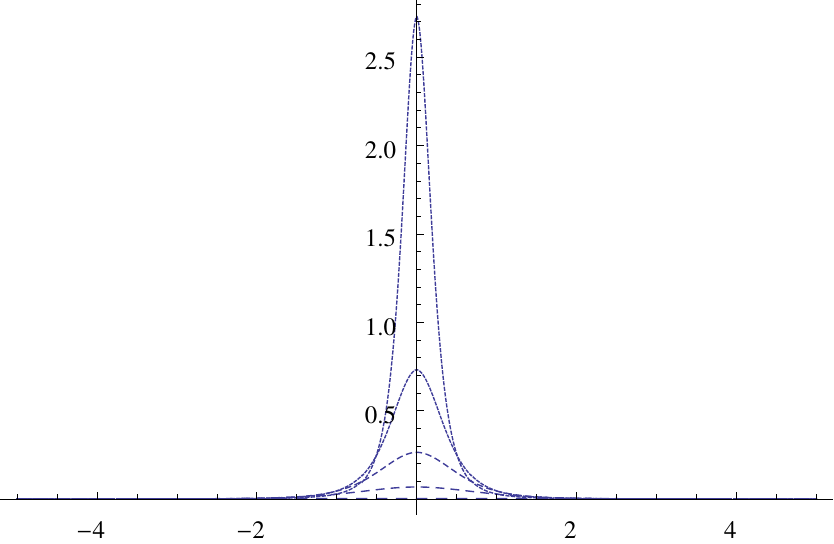}\caption{An S-vortex for various values of $\zeta$. $\xi=1$, $\zeta=-1.9,-1.45,-1,0.55,0.1$, $g=1$.}
\label{fig:Svortex}
\end{center}
\end{figure}

\begin{figure}[htbp]
\begin{center}
\includegraphics[width=8cm]{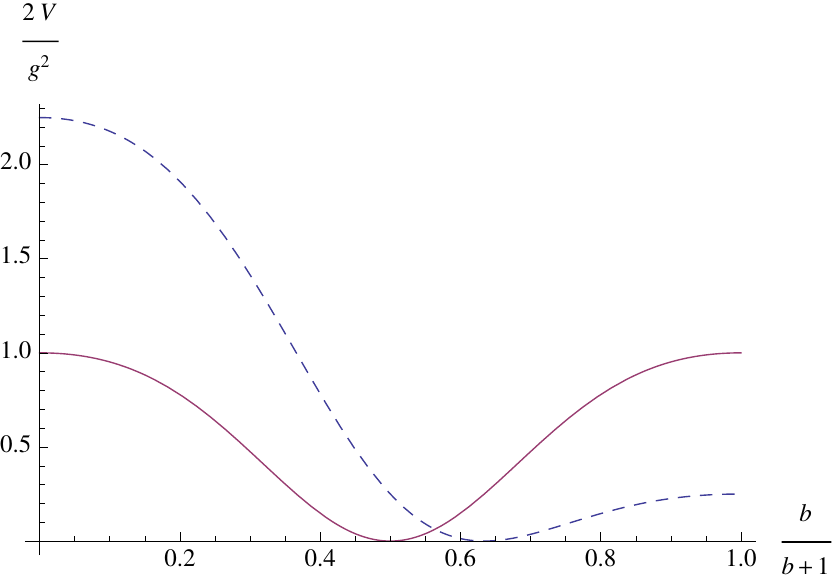}\caption{The potential as a function of $\beta=\frac{b}{1+b}$. This convenient combination takes into account the metric factor present in the kinetic term of $b$. Solid line: $\zeta=-1$. Dashed line: $\zeta=-1.5$.}
\label{fig:potential}
\end{center}
\end{figure}

Let us study a fundamental vortex in terms of $\zeta$. Fig.~\ref{fig:Svortex} shows the energy density profile  of an S-vortex as a function of the FI term $\zeta$. The vortex disappears, by dilution, in the limit $\zeta\rightarrow 0$, where the gauge symmetry defined on the S-patch is unbroken. In the other limit $\zeta\rightarrow-2\xi$, where the expectation value $b_{\infty}$ goes to infinity,  the vortex becomes a spike. 
There is no need for  additional work to study the N-vortex. From both the mass formula and the BPS equations, we see that an N-vortex is transformed into a physically equivalent  S-vortex by the formal replacements:
\beq
b \rightarrow 1/b, \quad \zeta \rightarrow -\zeta-2 \xi\,.
\eeq
A choice of parameters which gives a narrow S-vortex will then gives a wide N-vortex and vice versa. This can also easily understood if we plot the potential of Eq.~(\ref{eq:gauged}) as in Fig.~\ref{fig:potential}. If $\zeta$ is closer to $2 \xi$, the potential is higher around the S-pole than it is around the N-pole. A narrow S vortex is thus favored to minimize the potential energy with respect to the gradient energy (which tends to be higher for narrower configurations). Vice versa, for an N-vortex a wider configuration is favored. Notice that when $\zeta=-\xi$, the potential is symmetric and both vortices have the same width.  
We then have two independent effects which determine the size of vortices. A ``core'' effect, is related to the height of the potential which is different at the center of S and N vortices. A ``tail'' effect, instead, is related to the long distance, exponential fall-off of the vortex, and is the same for both S and N vortices. 
Analogously to the Abelian Higgs model case, we can estimate the typical sizes as follows
\beq
\lambda_{S}=\frac{1}{g \,\xi\, |b_{\infty}|}\,,\quad \lambda_{N}=\frac{1}{g\,\xi\,|b'_{\infty}|}=\frac{|b_{\infty}|}{g\,\xi}\,,
\label{eq:sizes}
\eeq
if we ignore the effects of the curvature of the target space on the vortex profile.
The exponential tail can then be checked analytically  expanding the master equation in terms of small fluctuations around the vacuum. We thus proceed expanding $\omega_{g}$ in the following way:
\beq
\omega=\left|\frac{b_{\infty}}{b_{0}(z)}\right|(1+\delta \omega+\mathcal O(\delta \omega^{2}))\,.
\label{eq:eq:expexp}
\eeq
The master equation (\ref{eq:master}) then linearizes:
\beq
\partial_{z}\partial_{\bar z}\delta \omega= g^{2 }\xi \frac{|b_{\infty}|^{2}}{(1+|b_{\infty}|^{2})^{2}}\delta \omega=-g^{2} \frac{\zeta(\zeta+2 \xi)}{4 \xi}\delta \omega\equiv \frac\lambda4 \delta \omega
\eeq
The solution of the equation above is the well-known modified Bessel function of the second kind:
\beq
\delta \omega(r) = K_{0}(\lambda r)\sim \frac{e^{-\lambda r}}{r}
\label{eq:exp}
\eeq
The exponential decay factor $\lambda$ corresponds to the mass of the $b$ field in the vacuum. Notice that $\lambda$ is invariant under change of coordinate, as it should be since it is a physical quantity.
%
%

\section{S-N Vortex System} 
If we send the gauge coupling  to zero we have to recover the un-gauged sigma model, which does not admit vortices. In fact, the fundamental S or N vortices become wider in this limit,  and eventually they vanish. 

However, there is something interesting which happens when we take the limit $g \rightarrow 0$ in the presence of a composite configuration of S and N vortices: we may recover lump solutions of the un-gauged NL$\sigma$M. We will study this phenomenon in this Section.

\subsection{Well separated S-N vortices}

Vortices are well separated when their typical widths  are much smaller than their separations:
\beq
\lambda_{S,N}\ll\Delta z_{i}\,.
\label{eq:strong}
\eeq
In this situation the moduli matrix (\ref{eq:modmatr}) obviously represents a  set of well separated $n_{N}$ N-vortices located at the poles of $b_{0}$ and $n_{S}$ S-vortices located at zeroes. We also refer to this situation as ``strong gauging'', since Eq.~(\ref{eq:strong}) can be always satisfied with a sufficiently large value of the gauge coupling (see Eq. (\ref{eq:sizes})).

\begin{figure}[htbp]
\begin{center}
\includegraphics[width=12cm]{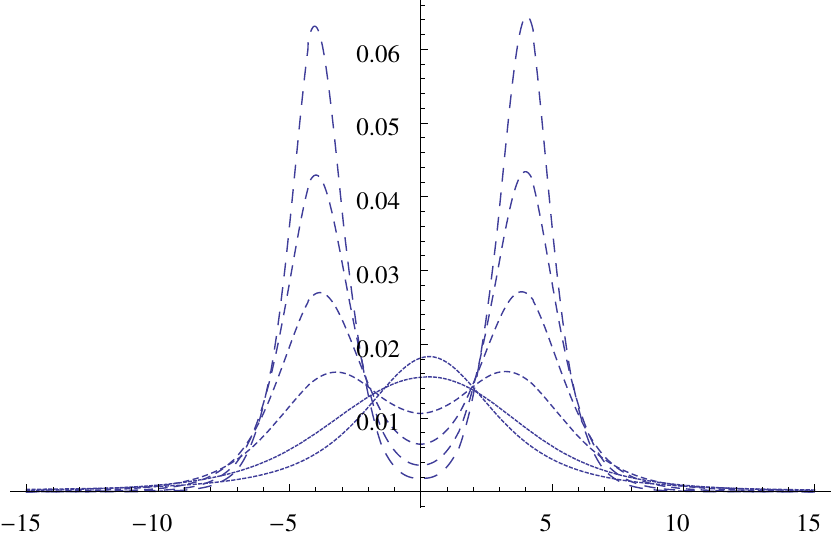}\caption{Deforming a lump into an N-S vortex system. $b_{\infty}=1$, $\xi=-\zeta=1$, $z_{S}=-z_{N}=-4$, $g=0,0.1,0.2,0.3,0.4,0.5$.}
\label{fig:composite}
\end{center}
\end{figure}

The simplest configuration is given by the following choice:
\beq
b_{0}(z)=b_{\infty}\frac{z-z_{S}}{z-z_{N}}\,.
\eeq
A numerical simulation in the case above is shown in Fig.~\ref{fig:composite} where an S and an N vortex are clearly identified when the gauge coupling is sufficiently large.

\subsection{``Gauged'' lumps}
Let us now consider the limit $g \rightarrow 0$. We want to recover the lump solutions in the original un-gauged NL$\sigma$M described by Eq.~(\ref{eq:ratmap}). Looking at the first line of Eq.~(\ref{eq:def}) and the definition of the moduli matrix $b_{0}$ in Eq.~(\ref{eq:modmatr}) we see that $b$ must reduce to a purely holomorphic function. Assuming that all the holomorphic factors are extracted in the ratio $b_{0}(z)$, we have
\beq
\omega\rightarrow 1& {\rm as} & g\rightarrow 0\,.
\label{eq:lumplim}
\eeq
Moreover, if  $b_{\infty}\neq0$,  the number of N and S vortices must be the same:
\beq
n_{S}=n_{N}=N\,.
\eeq
 In this situation the lump number defined in Eq.~(\ref{eq:lumpnumb}) is a well-defined integer.  If the gauge coupling is small enough (weak gauging regime), we expect to find solitonic solutions similar to lumps of the un-gauged  model at small distances. Nevertheless, at large distances this solitons will show their true nature of vortices falling off with an exponential factor. This behavior can be confirmed with numerical simulations. Again,  Fig.~\ref{fig:composite} shows how a lump, in the weak gauging regime, which is deformed into a composite state of S and N vortices once the  gauge coupling is strong enough.

We can also perform, to some extent,  an analytical analysis of the weak coupling limit, by expanding the non-holomorphic term $w$ and the master equation (\ref{eq:master})  as follows: 
\beq
w & = & 1+\delta w+ \mathcal O (g^{2}), \nonumber \\
\partial_{\bar z}\partial_{z} \delta \omega & = & -\frac{g^{2}}{4} \left(\xi\frac{-2|b_{0}|^{2}}{1+|b_{0}|^{2}}-\zeta \right)+\mathcal O (g^{2})\,.
\label{eq:lumpexp}
\eeq
 Let us study the equation above  in the proximity of  a  zero (pole) around $z=z_{0}$. The second term can  be approximated by  a constant in the vicinity of $z_{0}$:
\beq
\partial_{\bar z}\partial_{z} \delta \omega & = & \frac{g^{2}}{4}(\zeta+\xi\pm\xi)\,, \quad |z-z_{0}|\ll \Delta z_{i}\,,
\eeq
where the $\Delta z_{i}$ are the typical separations between zeros and poles. The plus and minus signs respectively apply to the case of zeroes and poles.
The equation above can then be easily solved for $\delta \omega$:
\beq
\delta \omega \sim   \frac{g^{2}}{4}(\zeta+\xi\pm\xi) |z-z_{0}|^{2}+ harmonics\,.
\eeq
where with  harmonics we mean solution of the homogenous equation $\partial_{\bar z}\partial_{z} \delta \omega  =0$\footnote{These are fluctuating solutions which are not important in the present argument.}.  Since the non-homogeneous term grows with the distance from $z_{0}$, the lump approximation is valid when:
\beq
|z-z_{0}|^{2}\ll|\Delta z_{i}|^{2}  \ll  \left(\frac{g^{2}}{4}(\zeta+\xi\pm\xi)\right)^{-1}\,.
\eeq
The condition above also tells us that a generic configuration of S and N vortices is approximated by a lump solution whenever the typical distances $|\Delta z_{i}|$ are small enough.
 Notice also that the approximation above is valid at short distances only. At sufficiently  large distances, in fact, we have to spot the true ``local'' nature of the  configuration for finite $g$.  Indeed,  the second term of Eq.~(\ref{eq:lumpexp}) vanishes at large distances, and we have to take into account terms linear in $\delta \omega$ in the expansion. We then recover Eq.~(\ref{eq:eq:expexp})
 which gives the correct exponential tail.

The metric on the moduli space of composite S and N vortices was also  previously studied in \cite{Baptista:2004rk,Baptista:2005bt,Baptista:2010rv}, where it was found that a  singularity develops  when  a couple  of  S-N vortices coincides.  In this Section we have shown that a configuration  of  two very close vortices (a pole and a zero) is indistinguishable from a lump  (see Eq.~(\ref{eq:posandsize}). The singularity discovered in Refs. \cite{Baptista:2004rk,Baptista:2005bt,Baptista:2010rv}  is then exactly identified as a small lump singularity. As well-known, small lump singularities can be  ``resolved'' by introducing an appropriate UV completion for the NL$\sigma$M. In the next Section we will consider this possibility explicitly.

\section{Linear Formulation}

\subsection{Model}
It is simple to guess the correct UV completion of the gauged $\mathbb CP^{1}$ NL$\sigma$M. First let us consider the un-gauged  NL$\sigma$M. It can be easily written as the strong gauge limit $e\rightarrow \infty$  of the so-called extended Abelian Higgs model with two flavors:
\beq
\mathcal L&  = & -\frac{1}{4 e^{2}} F_{e}^{\mu\nu}F_{e\mu\nu}+ |\nabla_{\mu} \phi_{1}|^{2}+|\nabla_{\mu} \phi_{2}|^{2}- \frac{e^{2}}{2}\left(|\phi_{1}|^{2}+|\phi_{2}|^{2}-\xi\right)^{2}\,.
\label{eq:exthiggs}
\eeq
As well-known, the model above admits semi-local vortices which  are a particular type of strings which admit size moduli \cite{Vachaspati:1991dz,Achucarro:1999it,Hindmarsh:1991jq,Hindmarsh:1992yy,Preskill:1992bf,Eto:2007yv}. They are indeed very similar to lump solutions of the underlying  $\mathbb CP^{1}$ NL$\sigma$M (to which they reduce in the $e\rightarrow \infty$ limit). However, while a lump becomes a singular spike in the zero size limit, a semi-local vortex remains regular and reduces to a traditional (sometimes called ``local'') Abrikosov-Nielesen-Olsen vortex, which has a typical size of order $1/e\sqrt\xi$. In this sense, the moduli space of semi-local vortices is considered to be a regularization of the singularities (small lump singularities) of the lump moduli space.

Semi-local vortices ultimately exist because  the vacuum manifold, in this case $S^{3}$, is simply connected
\beq
\pi_{1}(S^{3})=0\,,
\eeq
 while the second homotopy group of the moduli space of vacua $S^{3}/U(1)=\mathbb CP^{1}$  is non trivial \cite{Preskill:1992bf}
 \beq
 \pi_{2}(\mathbb CP^{1})=\mathbb Z\,.
 \eeq

Notice that the $SU(2)$ isometry of the $\mathbb CP^{1}$ NL$\sigma$M is now seen as a flavor symmetry which rotates the fields $\phi_{i}$. Moreover, we can identify the holomorphic coordinate  $b$  appearing in Eq. (\ref{eq:ungauged})  as:
\beq
b=\frac{\phi_{2}}{\phi_{1}}\,.
\eeq
It is then clear that to obtain the model (\ref{eq:gauged}) in the $e\rightarrow\infty$ we have to gauge a $U(1)$ flavor symmetry with the charge assignments given in Table 1.
\begin{table}[htdp]
\begin{center}
\begin{tabular}{c|cc}
$U(1)^{2}$ & $\phi_{1}$ & $\phi_{2}$ \\
\hline
$U(1)_{e}$ & 1 & 1\\
$U(1)_{g }$ &  1  & -1\\
\end{tabular}
\caption{Field content and charges}
\end{center}
\label{tab}
\end{table}%
The resulting model is then the following linear gauged model 
\beq
\mathcal L&  = & -\frac{1}{4 e^{2}} F_{e}^{\mu\nu}F_{e\mu\nu}-\frac{1}{4 g^{2}} F_{g}^{\mu\nu}F_{g\mu\nu}+ |\nabla_{\mu} \phi_{1}|^{2}+|\nabla_{\mu} \phi_{2}|^{2}+\nonumber \\
& - & \frac{e^{2}}{2}\left(|\phi_{1}|^{2}+|\phi_{2}|^{2}-\xi\right)^{2}-\frac{g^{2}}{2}\left(|\phi_{1}|^{2}-|\phi_{2}|^{2}-\chi\right)^{2},
\label{eq:gaugedextended}
\eeq
where $\chi$ is a  second FI term for the new gauge group $U(1)_{g}$ and  
\beq
\nabla_{\mu}\phi_{1}=(\partial_{\mu}-i A_{e\mu}-i A_{g\mu})\phi_{1},\quad \nabla_{\mu}\phi_{2}=(\partial_{\mu}-i A_{e\mu}+i A_{g\mu})\phi_{2}\,.
\eeq
By comparing the second potential term in the equation above with the potential term arising in the gauged NL$\sigma$M lagrangian (\ref{eq:gauged}) we also obtain the following relationship between the FI terms $\chi$ and $\zeta$
\beq
\chi=\zeta+\xi\,.
\eeq
The vacuum of the theory is then given by:
\beq
|\phi_{1\infty}|^{2}=\frac{\xi+\chi}{2},\quad |\phi_{2\infty}|^{2}=\frac{\xi-\chi}{2}\,.
\eeq
Both the $U(1)$ gauge groups are then spontaneously broken, we thus have for the first homotopy group:
\beq
\pi_{1}\left( \frac{U(1)_{e}\times U(1)_{g}}{\mathbb Z_{2}}\right)=\mathbb Z_{S}\times \mathbb Z_{N}\,.
\eeq
Notice that, due to the presence of a $\mathbb Z_{2}$ common factor, the smallest closed paths are obtained with  a $\pi$ rotation into the group $U(1)_{e}$ and either a $\pi$ or a $-\pi$ rotation into the group $U(1)_{g}$. These two paths represent the smallest elements of  $\mathbb Z_{S}$ and $ \mathbb Z_{N}$ respectively. The choice of notation is not a coincidence either: the two factors characterize two type of vortices which correspond to the  S and N vortices we already studied in the previous Sections in the $e \rightarrow \infty$ limit. The S and N vortices are a particular type  of what is called, in literature, fractional vortex. They were first discovered  in two component superconductors within  a Landau-Ginzburg model 
\cite{Babaev:2001hv,Babaev:2002ck}, which can be considered as a non-relativistic version of the model  (\ref{eq:gaugedextended}) where, in the case of superconductors, the group $U(1)_{g}$ is global\footnote{Different types of fractional vortices were also introduced in Ref. \cite{Collie:2009iz,Eto:2009bz}.}. This definition is due to the fact that the fundamental vortices carry only one half of the magnetic flux of both gauge groups (in particular of the group $U(1)_{e}$). For generic values of the gauge couplings, the S and N vortices do not feel static interactions, even if they form a coupled system. Non-static interactions are on the other hand non-trivial and can be studied in the moduli space approximation  \cite{Manton:1981mp}. However, when the gauge couplings coincide $e=g$ the S and N vortices completely decouple and do not interact at all, at least at the classical.

\subsection{Vortices: BPS equations and moduli matices}
As usual we can perform a Bogomol'nyi completion of the action 
\beq
\mathcal E& =&  \frac{1}{2e^{2}}\left[F_{e12}-e^{2}\left(|\phi_{1}|^{2}+|\phi_{2}|^{2}-\xi\right)\right]^{2}+ \frac{1}{2g^{2}}\left[F_{g12}-g^{2}\left(|\phi_{1}|^{2}-|\phi_{2}|^{2}-\chi\right)\right]^{2}+\nonumber\\[2mm]
& + &  4 |\nabla_{\bar z} \phi_{1}|^{2} +  4 |\nabla_{\bar z} \phi_{2}|^{2}  -\xi F_{e12} -\chi F_{g12}  +2\partial_{z}\partial_{\bar z}(|\phi_{1}|^{2}+|\phi_{2}|^{2})\,,
\label{eq:linearbogcompl}
\eeq
which leads to BPS equations \cite{Schroers:1995ns,Schroers:1996zy}:
\beq
F_{e12}&=&e^{2}\left(|\phi_{1}|^{2}+|\phi_{2}|^{2}-\xi\right) \nonumber \\
F_{g12}&=&g^{2}\left(|\phi_{1}|^{2}-|\phi_{2}|^{2}-\chi\right) \nonumber \\
\nabla_{\bar z} \phi_{1}& = & \nabla_{\bar z} \phi_{2}=0\,.
\eeq
The tension is then given by the following topological term:
\beq
T=2\pi \xi\,\nu_{e}+2\pi \chi\,\nu_{g}
\eeq
where $\nu_{e}$ and $\nu_{g}$ are respectively the winding numbers for the gauge groups $U(1)_{e}$ and $U(1)_{g}$. Notice that the last term in Eq.~(\ref{eq:linearbogcompl}) is a total derivative with vanishing  contribution.

The moduli matrix formalism is implemented as usual with the substitutions \cite{Eto:2008yi,Eto:2009bg}:
\beq
\phi_{1}(z,\bar z)  & = & s_{e}^{-1}(z,\bar z)s_{g}^{-1}(z,\bar z) \phi_{01}(z), \nonumber \\
\phi_{2}(z,\bar z)  & = & s_{e}^{-1}(z,\bar z)s_{g}(z,\bar z) \phi_{02}(z), \nonumber \\
A_{e\bar z} & = & -i \partial_{\bar z} \log s_{e}, \nonumber \\
A_{g\bar z} & = & -i \partial_{\bar z} \log s_{g}, \nonumber \\
\omega_{e}=|s_{e}|^{2}, & & \omega_{g}=|s_{g}|^{2}\,.
\eeq
 $\phi_{01}(z)$ and $\phi_{02}(z)$ are the moduli matrices for $\phi_{1}$ and $\phi_{2}$ respectively. The general prescription  to find the right boundary conditions on the moduli matrix was described in Ref. \cite{Eto:2008yi,Eto:2009bg} and gives in the present case 
\beq
\phi_{01}(z)& = & \sqrt\frac{\xi+\chi}{2}(z^{n_{N}}+p_{1} z^{n_{N}-1}+\dots +p_{n_{N}}) \nonumber \\
\phi_{02}(z) & = & \sqrt\frac{\xi-\chi}{2}(z^{n_{S}}+q_{1} z^{n_{S}-1}+\dots +q_{n_{S}}) \nonumber \\
\omega_{e}\rightarrow  |z|^{n_{N}+n_{S}}, & & \omega_{g}\rightarrow |z|^{n_{N}-n_{S}}\,,
\eeq
where $n_{S}$ and $n_{N}$ are the vortex numbers
\beq
\nu_{e}&=&  -\frac1{2\pi}\int dx^{2} F_{e12}=- \frac1\pi \int dx^{2} \partial_{z}\partial_{\bar z} \log \omega_{e}= \frac{n_{S}+n_{N}}{2}, \nonumber \\
\nu_{g}&=& -\frac1{2\pi}\int dx^{2} F_{g12}=- \frac1\pi \int dx^{2} \partial_{z}\partial_{\bar z} \log \omega_{g}= \frac{n_{S}-n_{N}}{2}\,.
\eeq
In the language of the NL$\sigma$M the integers above correspond to the (fractional) vortex and lump numbers defined in the previous section:
 \beq
 \nu_{g}&=&\nu_{v}\nonumber \\
 \nu_{e}&=&\nu_{l}\,.
 \eeq
Finally, let us write the BPS equations in terms of the moduli matrices:
\beq
\partial_{\bar z}\partial_{z} \log \omega_{e}& = & \frac{e^{2}}{4}\left(\omega_{e}^{-1}\omega_{g}^{-1}|\phi_{01}|^{2}+\omega_{e}^{-1}\omega_{g}|\phi_{02}|^{2} -\xi\right) ,\nonumber \\
\partial_{\bar z}\partial_{z} \log \omega_{g}& = & \frac{g^{2}}{4}\left(\omega_{e}^{-1}\omega_{g}^{-1}|\phi_{01}|^{2}-\omega_{e}^{-1}\omega_{g}|\phi_{02}|^{2} -\chi\right) .
\label{eq:linearbpseq}
\eeq

\subsection{Resolving the small lump singularity}
One can easily check that all the analysis of this Section correctly reduces to that done for the gauged $\mathbb CP^{1}$ NL$\sigma$M in the limit $e\rightarrow\infty$. Moreover, at finite $e$, but in the regime $e\gg g$, the results of the sections about the gauged NL$\sigma$M qualitatively apply to the linear case. We have just to substitute lumps with the very similar semi-local vortices in the discussions of the previous Sections. A semi-local vortex is thus split  into a couple of S and N vortices upon gauging of a $U(1)$ flavor symmetry.
  
However, the crucial difference is that small lump singularities are now removed. As we have seen in the previous section, these singularities arise when a couple of S and N vortices have coincident positions: there, a singular spiky lump develops. At finite $e$, however, this singularity is substituted by the insertion of a regular local vortex. To see this, let us consider the moduli matrices in the case of two coincident S and N vortices. For simplicity, we can  set $\chi=0$. Then we have
\beq
\phi_{01}=  \phi_{02} =   \frac{\xi}{2}(z-z_{0})
\eeq
and the BPS equations (\ref{eq:linearbpseq})  can be reduced to the following:
\beq
\omega_{g} & \equiv & 1\nonumber \\
\partial_{\bar z}\partial_{z} \log \omega_{e}& = & -\frac{e^{2}}{4}\left(2 \omega_{e}^{-1}|\phi_{01}|^{2} -\xi\right)= -\frac{e^{2}}{2}\left( \omega_{e}^{-1}|\phi_{01}|^{2} -\xi/2\right),
\eeq
 which represent the master equation for a standard, regular vortex for the $U(1)_{e}$ group.

\section{Regularization of the Semi-Local Vortex Metric}
As well-known, the effective theory on the world-sheet of  semi-local vortices (and similarly of lumps) has an infrared logarithmic divergence due to the slow polynomial decay of the fields \cite{Hindmarsh:1991jq,Vachaspati:1991dz,Eto:2007yv,Shifman:2006kd}. As usual for effective theories with at least four supercharges, the effective theory can be conveniently written in terms of a K\"ahler potential. In the case of a  semi-local vortex  the K\"ahler potential can be computed exactly at each order in a power expansion in terms of $1/(e|\rho|\sqrt\xi)$, where $|\rho|$ is the size modulus of the semi-local vortex \cite{Shifman:2011xc}. At the leading order in this expansion, the K\"ahler potential for a semi-local vortex in the Abelian extended Higgs model (\ref{eq:exthiggs}), is the following\footnote{High order corrections have been  calculated in \cite{Shifman:2011xc}.}
\beq
 K(|z-z_{0}|^{2},|\rho|^{2})=  2 \pi \xi |z-z_{0}|^{2} +\pi \xi |\rho|^{2}\ln \left(\frac{L^{2}}{|\rho|^{2}} \right)+\pi \xi |\rho|^{2}\,,
 \label{eq:semilockahler}
 \eeq
where we included the first term which describes the position $z_{0}$ of the vortex. In the expression above, the divergent logarithm has been regularized with the introduction of a large infrared cut-off $L$.

So far, various different approaches have been considered to consistently deal with these divergences. Two possibilities were  discussed in Ref. \cite{Shifman:2006kd}. One is to consider strings of finite length $L$. This would naturally cut-off the divergences as in (\ref{eq:semilockahler}). However, this approach obviously spoils the BPS nature of the string. A more convenient way to proceed is to introduce a twisted mass $m$ for the size modulus $\rho$. The logarithm is then cut off at the scale $L\sim1/m$. The clear advantage of this approach is that the massive deformation preserve the BPS nature of the vortex. However, as a drawback, this regularization is ultimately due to the fact that we really lift the moduli space once we give a mass to the size parameter $\rho$.   Another recent proposal takes advantage of the presence of the divergencies and of the possibility to eliminate them with an appropriate change of variables \cite{Shifman:2011xc}. The effective action is then exactly given by just this rescaled divergent terms.

In this work, we propose a new rather interesting regularization of the metric on the moduli space of semi-local vortices which preserves the BPS nature of the string and does not lift its moduli space. As a consequence  $\rho$ remains a genuinely massless  zero mode. This regularization is obtained by weakly gauging a flavor symmetry. While the regularization through a massive deformation corresponds to the inclusion F-terms, or potential terms, into the effective action,  the proposed regularization through weak gauging corresponds to a deformation of the K\"ahler potential. The example we studied in the previous Section is the explicit realization of this idea for the semi-local vortex in the extended Abelian Higgs model with two flavors, when we can gauge the relative phase between the two fields.

 Recalling the discussion of the previous Sections, it is straightforward to determine how the divergent  K\"ahler potential in (\ref{eq:semilockahler}) gets deformed, at least in two special limits. The first one is when the size of the vortex is small  $|\rho|^{2}\ll1/g^{2}\xi$ :
 \beq
 K(|z-z_{0}|^{2},|\rho|^{2})=  2 \pi \xi |z-z_{0}|^{2} +\pi \xi |\rho|^{2}\ln \left(\frac{1}{|\rho|^{2}g^{2}\xi} \right)+\pi \xi |\rho|^{2}\,.
 \label{eq:eff1}
 \eeq
The expression above can be easily guessed if we notice that after gauging the power law fall-off of the fields is cut off with an exponential behavior, at distances of order of the mass of the lightest particles in the bulk. In the limit of large size $|\rho|^{2}g^{2}\xi\gg1$, on the other hand, the semi-local vortex is split into two local vortices, and the K\"ahler potential reduces to that of two isolated Abelian vortices\footnote{We have set $\chi=0$ for simplicity, throughout this Section.}:
\beq
 K(|z-z_{N}|^{2},|z-z_{S}|^{2})=  \pi \xi |z-z_{N}|^{2} +\pi \xi |z-z_{S}|^{2}=  2 \pi \xi |z-z_{0}|^{2} +2 \pi \xi |\rho|^{2}\,,
 \label{eq:eff2}
 \eeq

The  K\"ahler potential at generic values of the coupling $g$ must then be calculated numerically using the expression
\beq
K(|z-z_{0}|^{2},|\rho|^{2})&=&   \int d^{2}x \left\{ \xi \ln {\omega_{e}}+\chi\ln\omega_{g}+\frac{1}{e^{2}}\partial_{i}\ln \omega_{e}\partial_{i}\ln \omega_{e}+\frac{1}{g^{2}}\partial_{i}\ln \omega_{g}\partial_{i}\ln \omega_{g}+ \right.\nonumber \\
&+ & \omega_{e}^{-1}\omega_{g}^{-1}|\phi_{01}|^{2}+\omega_{e}^{-1}\omega_{g}|\phi_{02}|^{2}\bigg\}\,,
\eeq
whose numerical evaluation must interpolates between the expressions (\ref{eq:eff1}) and (\ref{eq:eff2}).

\section{Generalizations}
In the case of the $\mathbb CP^{1}$ NL$\sigma$M, the $U(1)$ sub-group of the $SU(2)$ isometry is the maximal symmetry which can be gauged without breaking supersymmetry \cite{Bagger:1982fn}. This fact can be intuitively understood by recalling that gauging in supersymmetric theories generically reduces the complex dimension $D$ of a  target manifold to   $D-d$, where $d$ is the dimension of the  group being gauged\footnote{Gauging in a supersymmetric compatible way requires the introduction of a D-term potential, whose minimization gives generically $d$ real constraints, in addition to the $d$ degrees of freedom eaten by gauge invariance.}. At most, we can then gauge $D$ isometries.

The discussion of the previous sections can be generalized in the case of  NL$\sigma$Ms with higher dimensional target spaces $D>1$ in a variety of ways. Gauging in a different way the isometries of the target space will reveal different interesting connections between various type of vortices an lumps. In this Section we will have a brief look at some interesting possibilities.

\subsection{$\mathbb CP^{N-1}$ NL$\sigma$M}

The $\mathbb CP^{N-1}$ NL$\sigma$M has a $SU(N)$ isometry. As already explained, gauging of the full isometry will break supersymmetry \cite{Bagger:1982fn}. We can  however safely consider the gauging of a subgroup of dimension at most $N-1$. We will consider, as particular examples, gauging of Abelian and  non-Abelian subgroups.

\subsubsection*{Gauging of an Abelian isometry}
A simple generalization of the model discussed in Section \ref{sec:gauged} is the one given by weakly gauging a $U(1)^{d}$ subgroup with charge assignments given by Table 2.
\begin{table}[htdp]
\begin{center}
\begin{tabular}{c|ccccc}
$U(1)^{d}$ & $b_{1}$ & \dots &$b_{d}$ & \dots &  $b_{N-1}$ \\
\hline
$U(1)_{1}$ & -2 & 0 & 0& 0&0  \\
\vdots &  0  & $\ddots$ &0 &0 &0  \\
$U(1)_{d}$ &   0 & 0 & -2 &0 &0  \\
\end{tabular}
\caption{Field content and charges}
\end{center}
\label{tab:cpn}
\end{table}%
After gauging, the target space of  $\mathbb CP^{N-1}$ is effectively reduced to  $\mathbb CP^{N-1-d}$. When gauge couplings are sent to infinity, then, the model supports $\mathbb CP^{N-1-d}$ lumps. Moreover, at finite values of gauge couplings, the Abelian gauge symmetries are generically spontaneously broken, and the model will support abelian vortices too:
\beq
\pi_{1}(U(1)^{d})=\mathbb Z^{d}.
\eeq

Much in the same way of the $\mathbb CP^{1}$ case, a lump of the  un-gauged $\mathbb CP^{N-1}$ sigma model will thus split into a composite state of a semi-local vortex (which reduces to a $\mathbb CP^{N-1-d}$ lump in the infinite gauge coupling limit) plus $d$ Abelian vortices. In other words, $d$ size moduli of the original lump are transformed into an equivalent number of position moduli of Abelian vortices.

\subsubsection*{Gauging of a non-Abelian isometry}
Another interesting possibility is to gauge a non-Abelian $U(d)$ isometry, with $d^{2}\le N-1$. As shown in Table 3, we can arrange $d^{2}$ fields as $d$ fundamentals of $U(d)$, while the rest are singlets.
\begin{table}[htdp]
\begin{center}
\begin{tabular}{c|c|c}
 & $b_{1},\dots, b_{d^{2}}$ &  $b_{d^{2}+1},\dots,b_{N-1}$ \\
\hline
&&\\
$U(d)$ & $\Box$ &0  \\
\end{tabular}
\caption{Field content and charges}
\end{center}
\label{tab:nonab}
\end{table}%
Again, after gauging the target space effectively reduces to  $\mathbb CP^{N-1-d^{2}}$, which still supports lumps. Moreover, non-Abelian vortices are supported by the non-trivial homotopy:
\beq
\pi_{1}(U(d))=\mathbb Z\,.
\eeq
After gauging of a $U(d)$ isometry then, lumps of  $\mathbb CP^{N-1}$ sigma models reduce to a composite state of a semi-local vortex (similar to a $\mathbb CP^{N-1-d^{2}}$  lump) plus $d$ $U(d)$ non-Abelian vortices. The presence of $d$ vortices is due to the fact that non-Abelian vortices have $1/d$ charge with respect to an Abelian vortex. It can also be guessed by matching the dimensions of the moduli spaces. $d^{2}$ size moduli of the original lump are translated into  $d$ orientations of the $d$ non-Abelian vortices.

\subsection{Grassmannian NL$\sigma$M}

Grassmannian ($Gr_{N,\tilde N}$)  can be considered as a generalization of $\mathbb CP^{N-1}$ manifolds. $Gr_{N,\tilde N}$ is  the set of $N$-dimensional complex planes in a $N+\tilde N$ dimensional space. They can be algebraically defined as the set of $N\times(\tilde N+ N)$ matrices modulo an $SU(N)$ equivalence:
\beq
Gr_{N,\tilde N}=\left\{M_{N,\tilde N+ N}\,|\, M\sim G M,   \, G\in GL(\mathbb C, N)  \right\}\,.
\eeq
The complex dimension of the manifold is then $N\,\tilde N$. Grassmannian manifolds enjoy a $SU(N+\tilde N)$ isometry. If we write the matrix $M$ as follows:
\beq
M=(A|B)
\eeq
the isometry action is then:
\beq
 M\rightarrow M\,G_{SU(N+\tilde N)}\,.
\eeq
The matrices $A$ and $B$ have the role of homogeneous coordinates. Using the $GL(N,\mathbb C) $ action, and assuming that $A$ is invertible, we can always reduce $M$ to the following form:
\beq
M\sim ({\bf 1}|B')\,,
\eeq
where now the $B'$ fields are the independent $N\,\tilde N$ holomorphic coordinates of the grassmannian\footnote{Grassmannian manifolds are interesting in  quantum field theory since they give the Higgs branch  of supersymmetric $U(N)$ $\mathcal N=2$ QCD with $N+\tilde N$ fundamental flavors \cite{Seiberg:1994bz,Argyres:1996eh}.}. Grassmannian lumps where studied in connection to non-Abelian semi-local vortices in Ref. \cite{Eto:2007yv,Eto:2006db}. The dimension of the moduli space of a grassmannian lump is $N+\tilde N$. 

Assuming $\tilde N\le N$, we can choose to gauge a $U(\tilde N)$ subgroup of the $SU(N\times \tilde N)$ isometry acting  on the fields like
\beq
A\rightarrow A\,e^{-i \theta},\quad B\rightarrow B\, G_{SU(\tilde N)}\,e^{i \theta}\,.
\eeq
 The original $Gr_{N,\tilde N}$ manifold is then reduced to a gauged $Gr_{\tilde N,N-\tilde N}$ manifold. The gauged sigma model will then support non-Abelian $U(\tilde N)$ semi-local vortices, whose moduli space dimension is $\tilde N+(N-\tilde N)=N$. However, the original grassmannian lump will split into a composite state of a $U(\tilde N)$ semi-local vortex (which reduces to a $Gr_{\tilde N,N-\tilde N}$ lump in the infinite gauge coupling limit) plus an additional $U(\tilde N)$ local vortex.

\section{Summary and Conclusions}

In this paper we have explicitly constructed vortices in the $U(1)$ gauged  $\mathbb CP^{1}$ non-linear sigma model, and studied them numerically. We identified the configuration given by the superposition of N and S vortices as a two-dimensional instanton (lump) of the original un-gauged sigma model. In this way we identified a general mechanism to decompose  instantons into smaller constituents by gauging  an isometry of the target space of a non-linear sigma model. 

We can perform a similar analysis in the linear case of the Abelian extended Higgs model with two flavors where a $U(1)$ flavor symmetry is also gauged. This model can be seen as an UV completion of the gauged  $\mathbb CP^{1}$ sigma model. The extended Higgs model is well-known to contain semi-local vortices \cite{Vachaspati:1991dz,Achucarro:1999it}. In this context we identified the gauging of flavor symmetries as a general mechanism to regularize the metric of semi-local vortices. Contrarily to other known methods, the one we propose does not lift any moduli space parameter. 

We also briefly discussed how the above ideas can be generalized to the case of non-linear sigma models with higher dimensional target spaces. In general, lumps are split into component objects which can be identified as being vortices, non-Abelian vortices or semi-local vortices, depending on the chosen isometry one decides to gauge.

%

It is tantalizing to try to extend these ideas to the four dimensional case.  An interesting direct connection between two dimensional lumps and four dimensional instantons has been found in Ref. \cite{Eto:2004rz} through non-Abelian vortices \cite{Isozumi:2004vg,Eto:2005yh,Eto:2006pg,Hanany:2003hp,Auzzi:2003fs,Shifman:2004dr,Hanany:2004ea,Shifman:2007ce,Tong:2005un,Tong:2008qd}. Precisely,  instantons living in the un-broken phase of four dimensional non-Abelian $SU(N)$ gauge theories are confined on the two dimensional string world-sheet of non-Abelian vortices once one enters a broken Higgs phase. The effective theory of non-Abelian vortices is described by  a two dimensional $\mathbb CP^{N-1}$ non-linear sigma model and  instantons are then confined as lumps of the effective theory. Due to the results of the present paper, it is natural to expect that gauging of $U(1)$ flavor symmetries in the four dimensional theory may lead to instantons splitting into meron like objects. A situation where this set-up can be realized in nature is for example
the color-flavor-locked superconducting phase in high density QCD
\cite{Alford:2007xm}, where non-Abelian vortices
\cite{Balachandran:2005ev,Eto:2009kg}
admit $\mathbb CP^{2}$ moduli in the world-sheets
\cite{Nakano:2007dr,Eto:2009bh}
and the role of the additional $U(1)$ gauge symmetry is played by the
electromagnetic interactions. {Fractional instantons, or merons, on the world-sheet of non-Abelian strings  have already been considered in Ref. \cite{Auzzi:2009yw} in the different context of $\mathcal N=1^{*}$ $SU(N)$ gauge theories, where non-Abelian vortices are described by a  non-supersymmetric, un-gauged $O(3)$ sigma model \cite{Markov:2004mj}. In this case merons are identified as global vortices instead of our local vortices. Moreover, the setup is completely non supersymmetric.} We strongly believe that future research along this line  may lead to new crucial understandings of the vacuum structure of strongly coupled gauge theories.

\section*{Acknowledgments}
\addcontentsline{toc}{section}{Acknowledgments}
The work of MN is supported in part by 
Grant-in Aid for Scientific Research (No. 23740198) 
and by the ``Topological Quantum Phenomena'' 
Grant-in Aid for Scientific Research 
on Innovative Areas (No. 23103515)  
from the Ministry of Education, Culture, Sports, Science and Technology 
(MEXT) of Japan.\\
The work of WV is supported by the DOE grant DE-FG02-94ER40823.

\newpage

\appendix

\section{Superfield Formalism}
Let us explicitly derive the Bogomol'nyi equations for a general $U(1)$ gauged $\mathcal N=2$ non-linear sigma-model. Using an $\mathcal N=1$ superfield formalism, and restricting to the relevant bosonic content, the action of the model is given in terms of a K\"ahler potential and a gauge kinetic term  \cite{Bagger:1982fn}:
\beq
\mathcal L_{K}&=&\int d^{4} \theta \left\{K(e^{-V}|\Phi_{I}|^{2})+\zeta V\right\}+ \frac{1}{4g^{2}}\left(\int d^{2} \theta W^{\alpha} W_{\alpha}+c.c. \right) \nonumber \\
 W^{\alpha}&=&-\frac14D\bar D D_{\alpha} V\,,
\eeq
where the $\Phi_{I}$ are charged chiral superfields and and $V$ is the vector superfield.
The bosonic part is then the following:
\beq
\mathcal L_{K} &=&-\frac{1}{4 g^{2}} F_{g}^{\mu\nu}F_{g\mu\nu}+ g_{I\bar J}\nabla_{\mu}\phi^{I}\bar \nabla^{\mu}\phi^{\bar J}+\frac{g^{2}}{2}\left( \partial_{|\phi_{I}|^{2}}K  (|\phi_{I}|^{2})|\phi_{I}|^{2}-\zeta  \right)^{2}\,,\nonumber \\
g_{I\bar J} & = &\partial_{\phi_{I}\bar \phi_{\bar J}}K  (|\phi_{I}|^{2})
\eeq
One can then employ a Bogomol'nyi completion:
\beq
\mathcal E_{K}& =&  \frac{1}{2g^{2}}\left[F_{g12}-g^{2}\left(\partial_{|\phi_{I}|^{2}}K  (|\phi_{I}|^{2})|\phi_{I}|^{2}-\zeta \right)\right]^{2} + 4 \, g_{I\bar J}\nabla_{z}\phi^{I}\bar \nabla_{\bar z}\bar \phi^{\bar J}+\nonumber\\[2mm]
& - & \zeta F_{g12}  + \epsilon_{ij}\partial_{i}\mathcal N_{j}. \nonumber \\
\nonumber \\
\mathcal N_{j}& =   &     \frac{i}{2}\partial_{|\phi_{I}|^{2}}K  (|\phi_{I}|^{2})  \,\phi^{I}\nabla_{j}\bar \phi^{\bar I}+c.c.\,.
\label{eq:bogcomplgeneral}
\eeq
As usual, vortex equations are given by imposing the vanishing of the squares in the first line of the expression above:
\begin{eqnarray}
F_{g12} & = & g^{2}\left(\partial_{|\phi_{I}|^{2}}K  (|\phi_{I}|^{2})|\phi_{I}|^{2}-\zeta \right)\nonumber \\
\nabla_{g \bar z} \phi_{I}& = & 0 \,,
\label{eq:boggeneral}
\end{eqnarray}
where the second follows since the metric $g_{IJ}$ is assumed to be positive definite.

\bibliography{Bibliographysmart-1}
\bibliographystyle{nb}

\end{document}